\documentclass[preprint,amssymb,showpacs,superscriptaddress,floatfix]{revtex4-2}
\usepackage{graphicx}
\usepackage{dcolumn}
\usepackage{bm}
\usepackage[utf8]{inputenc}
\usepackage[T1]{fontenc}
\usepackage{mathptmx}
\usepackage{etoolbox}
\usepackage{float}
\usepackage{booktabs}
\usepackage{amsmath,amssymb}
\usepackage[dvipsnames]{xcolor}
\usepackage{subcaption}
\usepackage{mathtools}
\usepackage[normalem]{ulem}

\makeatletter
\def\@email#1#2{
 \endgroup
 \patchcmd{\titleblock@produce}
  {\frontmatter@RRAPformat}
  {\frontmatter@RRAPformat{\produce@RRAP{*#1\href{mailto:#2}{#2}}}\frontmatter@RRAPformat}
  {}{}
}
\makeatother

\usepackage{color}

\usepackage[utf8]{inputenc}
\usepackage[T1]{fontenc}

\begin{document}

\title{Engineering Synaptic Dynamics in Ag-Modified TaO$_x$ Memristive Devices}

\author{R. Leal Martir$^{1,2}$}
\author{W. Quiñonez$^{3}$}
\author{A.J.T. van der Ree $^{4,5}$}
\author{M. H. Aguirre $^{6, 7, 8}$}
\author{G. Palasantzas $^{4,5}$}
\author{M. J. Sánchez $^{1,2}$}
\author{D. Rubi$^{3,*}$}

\affiliation{$^{1}$Instituto de Nanociencia y Nanotecnología (INN), CONICET-CNEA, nodo Bariloche, 8400 San Carlos de Bariloche, Río Negro, Argentina.\\
$^{2}$Centro Atómico Bariloche, Instituto Balseiro (UNCuyo), CONICET, 8400 San Carlos de Bariloche, Río Negro, Argentina.\\ $^{3}$Laboratorio de Ablación Láser (INN-CONICET-CNEA), Centro Atómico Constituyentes, Gral. Paz 1499, San Martín, Argentina. \\ $^{4}$Zernike Institute for Advanced Materials, University of Groningen, 9747 AG Groningen, The Netherlands.  \\ $^{5}$CogniGron Center, University of Groningen, 9747 AG Groningen, The Netherlands. \\ $^{6}$Dept. de Física de la Materia Condensada, Universidad de Zaragoza, Pedro Cerbuna, 12, 50009 Zaragoza, Spain. \\ $^{7}$INMA—Instituto de Nanociencia y Materiales de Aragón, CSIC—Universidad de Zaragoza, Mariano Esquillor s/n, 50018 Zaragoza, Spain. \\ $^{8}$Laboratorio de Microscopías Avanzadas, Universidad de Zaragoza, Mariano Esquillor s/n, 50018 Zaragoza, Spain.
}

\email{diego.rubi@gmail.com}

\begin{abstract}

Engineering not only the magnitude but also the dynamics of synaptic weight updates is an important challenge for memristive neuromorphic hardware. Here, we show that Ag modification of TaO$_x$ memristors introduces an electrically selectable degree of freedom in synaptic depression through the interplay between Ag-related and oxygen-vacancy dynamics. Reference devices exhibit conventional bipolar resistive switching and rapidly saturating conductance updates, whereas Ag-modified devices display highly symmetric table-with-legs hysteresis loops, metastable intermediate resistance states, and a strongly non-monotonic dependence of depression dynamics on programming amplitude. By varying only the pulse amplitude, the same physical device can be driven among three distinct depression regimes: a gradual sigmoidal response at low stimulus, an abrupt update concentrated within a few pulses at intermediate stimulus, and a broadly distributed conductance evolution extending over more than one hundred pulses at higher stimulus.

To rationalize this behavior, we introduce a minimal coupled-state model in which a vacancy-related switching variable interacts with a slower Ag-related internal degree of freedom. The model reproduces the gradual--abrupt--gradual crossover using two individually monotonic field-activated processes, supporting a scenario in which Ag-related configurations modulate vacancy-mediated switching kinetics.

The functional relevance of this tunability is evaluated using a memristor-based multilayer perceptron in which synthetic depression trajectories reproducing the three experimentally observed regimes are incorporated into the synaptic update rule. For MNIST classification, the abrupt, sigmoidal, and slowly evolving responses yield accuracies of $\approx$ 50\%, 85\%, and 88\%, respectively, while the gradual regime reaches $\approx$ 72\% for Fashion-MNIST. The abrupt response also exhibits substantially larger run-to-run variability, highlighting the functional penalty associated with coarse conductance updates.

These results show that coupling Ag-related and oxygen-vacancy dynamics can transform synaptic depression from a fixed device characteristic into an electrically programmable property, enabling a single memristive synapse to access functionally distinct learning regimes without modifying its physical architecture.

\end{abstract}

\maketitle

\section{INTRODUCTION}

Engineering the dynamics of synaptic weight updates is emerging as one of the key challenges in neuromorphic hardware. While remarkable progress has been achieved in the development of software-based artificial neural networks, their implementation on conventional von Neumann architectures remains fundamentally limited by the physical separation between memory and processing units, resulting in excessive data transfer, latency, and energy consumption. Neuromorphic computing addresses this bottleneck by physically co-localizing memory and computation, enabling massive parallel information processing inspired by biological neural systems \cite{Strukov2008,Yang2013,Prezioso2015}.

Among the different hardware platforms currently under development, memristive devices are considered one of the most promising candidates for artificial synapses owing to their simple metal--insulator--metal architecture, nanoscale scalability, low operating energy, non-volatility, and analog conductance modulation \cite{Yang2013,Li2018}. Consequently, extensive efforts have been devoted to improving their synaptic performance through materials engineering, interface design, and optimized programming protocols. Parameters such as conductance linearity, dynamic range, symmetry, retention, endurance, cycle-to-cycle variability, and the number of accessible conductance states have become standard figures of merit for evaluating neuromorphic devices \cite{Yu2020,IelminiWaser_2016,Li2018, quinonez_2026}.

Nevertheless, most reported memristive synapses exhibit essentially fixed potentiation and depression characteristics once the device architecture has been established \cite{zhu_2024, covi_2015, sudheer_2023}. Although pulse amplitude, width, or number are routinely adjusted to control the magnitude of conductance updates \cite{diao_2025, gao_2023}, the functional evolution of the synaptic response generally remains predetermined by the device itself. This limitation contrasts with biological learning processes, where different learning stages rely on distinct adaptation dynamics, and with modern optimization strategies, in which the effective learning rate evolves continuously during training.

Recent studies increasingly indicate that the temporal evolution of synaptic weight updates may be as important as the number of accessible conductance states \cite{milano_2022, zhou_2022}. Synaptic update dynamics are not necessarily universally optimal across learning tasks. The effective weight resolution, nonlinearity, and distribution of conductance changes imposed by a given potentiation/depression trajectory may interact differently with the statistics of the dataset, the network architecture, and the optimization landscape. Consequently, a trajectory that is advantageous for one task may be suboptimal for another, suggesting electrical selection of distinct update dynamics within the same physical device could provide an additional degree of hardware adaptability. Rather than relying on a single fixed synaptic response, such tunability could allow the update characteristics of the device to be matched to the requirements of a given task or dataset. Despite its potential importance, this concept has received comparatively little attention in oxide-based memristive systems.

Among oxide memristors, TaO$_x$ has become one of the reference materials because of its excellent CMOS compatibility, reproducible resistive switching, and vacancy-mediated transport mechanisms \cite{Lee2011, park_2015}. Different approaches have been proposed to tailor its electrical behavior, including interface engineering \cite{LealMartir_2026, zhu_2017}, multilayer oxides \cite{yang_2012, lee_2011}, metallic doping \cite{zhu_2025, chen_2026, song_2023}, and the incorporation of metallic nanoparticles \cite{Spring2020,Ning_2021, jana_2025}. In particular, noble-metal nanoparticles have been shown to modify local electric-field distributions, defect chemistry, filament nucleation, and switching uniformity, resulting in improved analog conductance modulation and enhanced device reproducibility in several oxide systems \cite{yang_2014,Gao_2017, liu_2010, xiaobing_2018, yue_2023, meng_2020}. However, previous studies have primarily focused on optimizing conventional switching characteristics, whereas the possibility of engineering fundamentally different synaptic update dynamics through electrical programming has remained largely unexplored.

Here we demonstrate that Ag nanoparticle (AgNP) incorporation enables electrically programmable synaptic dynamics in TaO$_x$ memristive devices. By varying the programming pulse amplitude, the same device can be driven among three characteristic depression regimes, ranging from gradual sigmoidal behavior to abrupt saturation, and finally, to distributed non-saturating conductance evolution. This functionality is accompanied by highly symmetric table-with-legs hysteresis switching loops and metastable intermediate resistance states.

In the pristine device, Ag is detected near the upper Pt/TaO$_x$ interface, while microscopy does not reveal extended metallic Ag clusters within the oxide. Upon electrical stressing, the evolution of the electrical response is consistent with the incorporation of Ag-related species into the oxide and a concomitant modification of vacancy-mediated switching. This behavior contrasts with our previous study of deliberately grown Pt/Ta$_2$O$_5$/TaO$_2$/Pt bilayers, where AgNP incorporation selectively suppressed one of two pre-existing interfacial switching channels and stabilized a single switching mode \cite{LealMartir_2026}. The different response observed here highlights the strong dependence of Ag-induced functionality on the underlying oxide stoichiometry and defect landscape. Our results therefore show that engineering coupled Ag-related and vacancy dynamics provides a route to expand the functional capabilities of oxide memristive synapses beyond conventional conductance optimization.

\section{Methods}

A 50 nm TaO$_x$ thin film was deposited on a Si/Pt substrate by Pulsed Laser Deposition (PLD) at room temperature under an oxygen pressure of 0.1 mbar. After film growth, Ag NPs were deposited ex-situ by plasma sputtering using a home-modified Mantis Nanogen 50 unit \cite{Brink_2014, vanderRee_2024}. The deposition parameters were determined from a TEM grid onto which NPs were simultaneously deposited, using a Helios G4 scanning electron microscope operated at 18 kV in HAADF mode. The nanoparticles exhibit an average size of (9 ± 2) nm and a loading amount, quantified as surface coverage, of 6\%. Their dispersibility was assessed through the mean nearest-neighbor distance, yielding (20 ± 7) nm. During deposition, part of the sample was intentionally masked, leaving a region of the film free of AgNPs to be used as a control. Top Pt electrodes were subsequently fabricated by a combination of sputtering and optical lithography. Both regions, with and without NPs, underwent identical processing conditions, resulting in two types of devices on the same film: one containing NPs embedded within the Pt contact and one without NPs.

Scanning transmission electron microscopy with a high-angle annular dark-field detector (STEM-HAADF) was carried out using a Cs-probe-corrected Titan 60-300 keV (Thermo Fisher Scientific) operated at 300 kV and equipped with an AZtec EDS detector from Oxford Instruments, with an estimated error in chemical quantification of approximately 2\% \cite{Fer_2020}. Lamellae were prepared by focused ion beam milling in a Helios NanoLab 650 Dual Beam system using Ga$^+$ ions with acceleration voltages ranging from 5 to 30 kV. 

Electrical measurements were performed at room temperature using a dual-channel source-measure unit (Keithley 2636B, Tektronix) connected to a probe station. The bottom Pt electrode was grounded throughout all experiments, while electrical stimuli were applied to the top Pt electrode. The electrical response was characterized using a pulse-based measurement protocol. Current--voltage ($I$--$V$) characteristics were reconstructed from a sequence of discrete programming voltage pulses rather than continuous voltage sweeps. Unless otherwise stated, the programming voltage was applied according to the sequence
$0 \rightarrow +V_{\mathrm{max}} \rightarrow -V_{\mathrm{max}} \rightarrow 0$,
or to an equivalent asymmetric protocol, as specified for each experiment. The current was measured during the application of each programming pulse after a short settling time.


The non-volatile resistance state was monitored through hysteresis switching loops (HSLs) \cite{Fer_2020}. Immediately after every programming pulse, a non-interfering read pulse of $V_{\mathrm{read}}=100$~mV was applied and the remanent resistance ($R_{\mathrm{REM}}$) was calculated as $
R_{\mathrm{REM}}=\frac{V_{\mathrm{read}}}{I_{\mathrm{read}}},
$ where $I_{\mathrm{read}}$ is the current measured during the read pulse. This write--read protocol enabled the reconstruction of the remanent resistance as a function of the previously applied programming voltage while minimizing read-induced perturbations. The HSL representation provides direct access to the evolution of the non-volatile resistance state independently of the dynamic current flowing during programming, allowing the identification of intermediate metastable resistance states and complementary switching processes \cite{roz_2010}. 

Synaptic potentiation and depression experiments were performed by applying trains of identical voltage pulses while monitoring the remanent resistance after each pulse using the same 100~mV read voltage. The pulse amplitude and polarity were varied depending on the experiment, whereas the pulse width was kept constant (100 ms). Device conductance (G) was calculated as $
G=\frac{1}{R_{\mathrm{REM}}},
$ and used as the synaptic weight throughout the manuscript.

The device yield was 75$\%$ (12 out of 16 devices) for the reference
structures and 33$\%$ (5 out of 15 devices) for devices incorporating
AgNPs. The lower yield of the Ag-modified structures was primarily
associated with short-circuited devices, indicating that incorporation
of the metallic nanoparticles introduces additional fabrication
constraints. The AgNP incorporation process was not specifically
optimized for device yield, suggesting that further optimization of nanoparticle density and incorporation conditions may improve device yield.

\section{Experimental Characterization and electrical response}

\begin{figure}[H] 
\centering
\includegraphics[width=1\linewidth]{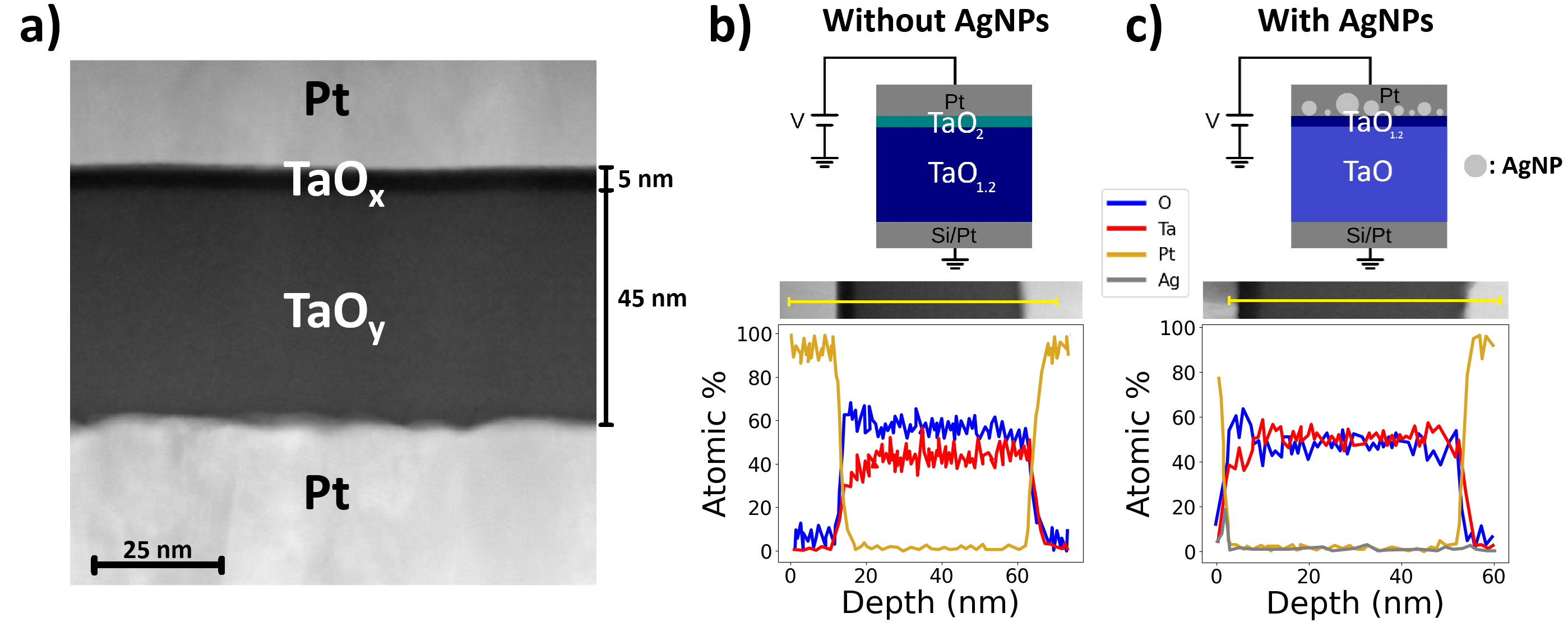}
\caption{Structural characterization of TaO$_x$ memristive devices with and without AgNPs. (a) Cross-sectional STEM-HAADF image of a pristine device containing AgNPs embedded within the top Pt electrode. The TaO$_x$ film consists of a thin, more oxidized upper layer ($\sim$5~nm) and a thicker, more reduced bottom layer ($\sim$45~nm). (b), (c) EDS line scans acquired across devices fabricated without and with AgNPs, respectively, showing the spatial distributions of Ta, O, Pt, and Ag. In the reference device, the upper and lower TaO$_x$ regions exhibit compositions close to TaO$_2$ and TaO$_{1.2}$, respectively, whereas in the Ag-containing device the oxide is overall more reduced, with compositions approaching TaO$_{1.2}$ in the upper region and TaO in the lower region. Although individual AgNPs are not resolved in the STEM image, the EDS profile of the Ag-containing device reveals a distinct Ag signal near the upper Pt/TaO$_x$ interface.}
\label{Fig1}
\end{figure}

FIG. \ref{Fig1}(a) shows a STEM-HAADF cross-sectional image of a pristine device containing AgNPs in its virgin state. The TaO$_x$ thin film is comprised of two different layers: a thin ($\sim$ 5 nm thick) more oxidized top layer, and a thicker ($\sim$ 45 nm thick) sub-oxidized bottom layer. The presence of the former is attributed to oxygen uptake upon exposure of the film to ambient conditions after growth and prior to Ag nanoparticle deposition.

EDS scans for pristine devices fabricated without and with AgNPs are shown in FIGs. \ref{Fig1}(b) and (c), respectively. For the device without AgNPs, the composition of the upper layer is close to TaO$_2$, whereas the bottom layer is more reduced, with a composition near TaO$_{1.2}$. The device containing AgNPs appears overall more reduced than the device without AgNPs, with the stoichiometry of the top layer approaching TaO$_{1.2}$ and that of the bottom layer nearing TaO. Both layers are fully amorphous.

The AgNPs are not directly visible in the STEM images, likely due to the predominance of the heavier Ta (Z = 73) and Pt (Z = 78) matrix in the STEM signal, which severely limits the detectability of the lighter Ag species (Z = 47). However, their presence was confirmed by EDS. Line scans of the device containing AgNPs, shown in FIG. \ref{Fig1}(c), reveal a distinct Ag signal near the Pt/TaO$_x$ interface that is absent in devices without AgNPs.

\begin{figure}[H] 
\centering
\includegraphics[width=0.9\linewidth]{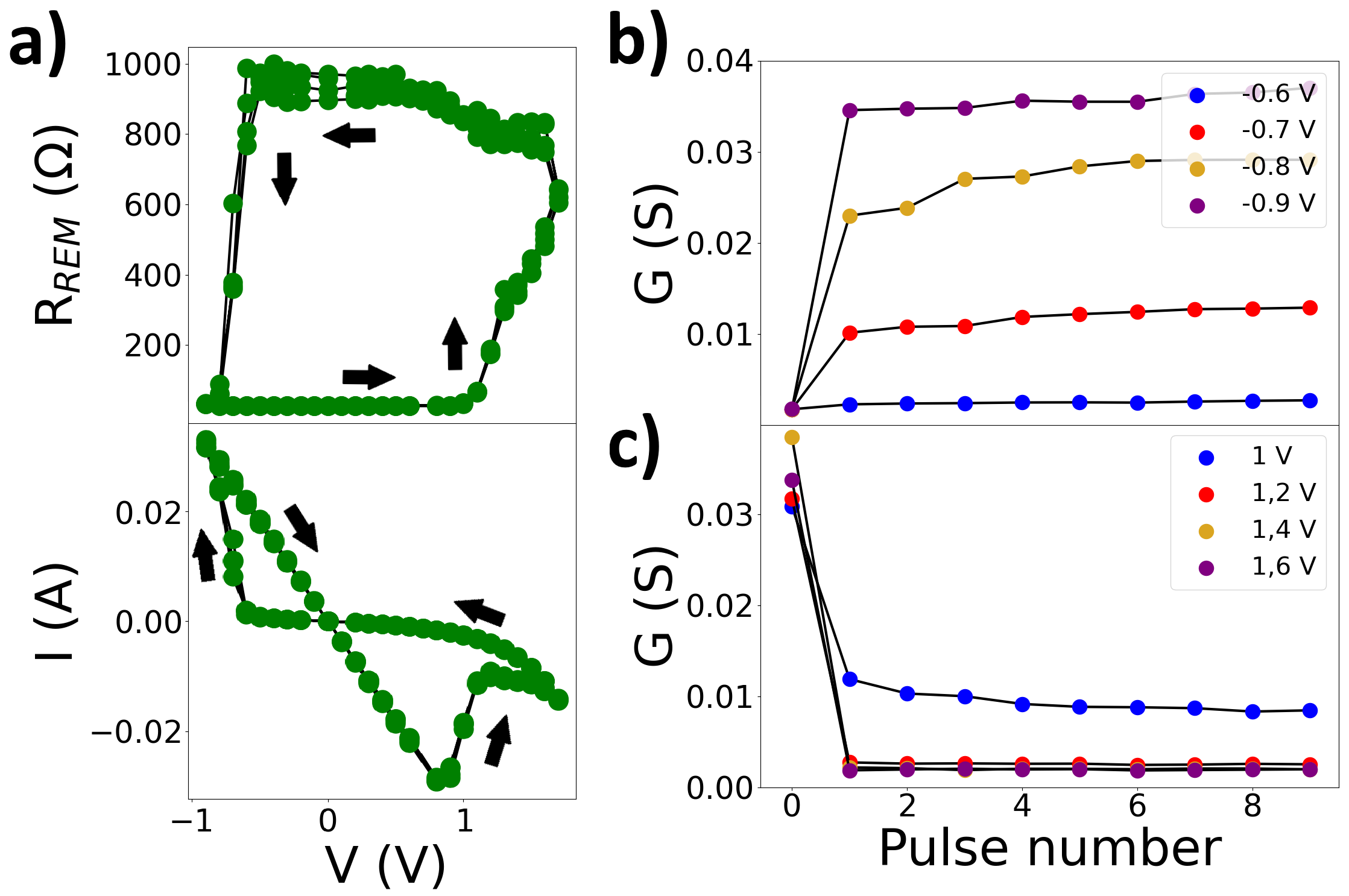}
\caption{Electrical characterization of reference TaO$_x$ memristive devices without AgNPs. (a) Remanent resistance, $R_{\mathrm{REM}}$ (top), and corresponding current--voltage ($I$--$V$) characteristics (bottom) measured after electroforming. The devices exhibit a conventional counter-clockwise (CCW) bipolar hysteresis switching loop, with low- and high-resistance states of approximately 40~$\Omega$ and 900~$\Omega$, respectively. The sharp SET transition is consistent with filamentary switching involving the redistribution of oxygen vacancies. (b) Potentiation curves measured under different programming pulse amplitudes, showing a rapid conductance increase followed by early saturation after only a few pulses. (c) Depression curves measured under different programming amplitudes, displaying a similarly fast conductance decrease and strongly saturating RESET dynamics for pulse amplitudes above approximately 1.2~V.}
\label{Fig2}
\end{figure}

FIG. \ref{Fig2} shows the electrical characterization of reference devices without AgNPs. The devices start in a virgin, high resistance state. Electroforming was achieved by repeatedly applying negative pulses in the range of -1 to -1.5 V, gradually lowering the device resistance until the operational range is reached.

The electroformed devices (FIG. \ref{Fig2}(a)) exhibit a counter-clockwise (CCW) hysteresis switching loop, characteristic of bipolar resistive switching dominated by a single active metal-oxide interface. The low- and high-resistance states are approximately 40~$\Omega$ and 900~$\Omega$, respectively, and the corresponding $I$--$V$ characteristics display the known pinched hysteresis typical of memristive devices. The sharp transition from the high- to the low-resistance state (SET process) is consistent with a filamentary switching mechanism involving the redistribution of mobile, positively charged oxygen vacancies, in agreement with previous reports on TaO$_x$-based systems \cite{LealMartir_2026}.

The reference devices exhibit conventional potentiation characteristics (FIG. \ref{Fig2}(b)). Increasing the programming voltage progressively increases the attainable conductance, while the conductance rapidly reaches its saturation value within the first few programming pulses. Only minor conductance variations are observed afterwards, indicating that the SET process is dominated by the rapid formation of a stable conductive path. The corresponding depression behavior is shown in FIG. \ref{Fig2}(c). For programming amplitudes above 1.2~V, the conductance decreases abruptly after the first pulse and rapidly reaches a nearly constant value, characteristic of a highly nonlinear low- to high resistance change (RESET) process. Lower programming voltages (1.0~V) produce a slightly more gradual conductance decrease, although the overall depression behavior remains rapidly saturating.

\begin{figure}[H] 
\centering
\includegraphics[width=1\linewidth]{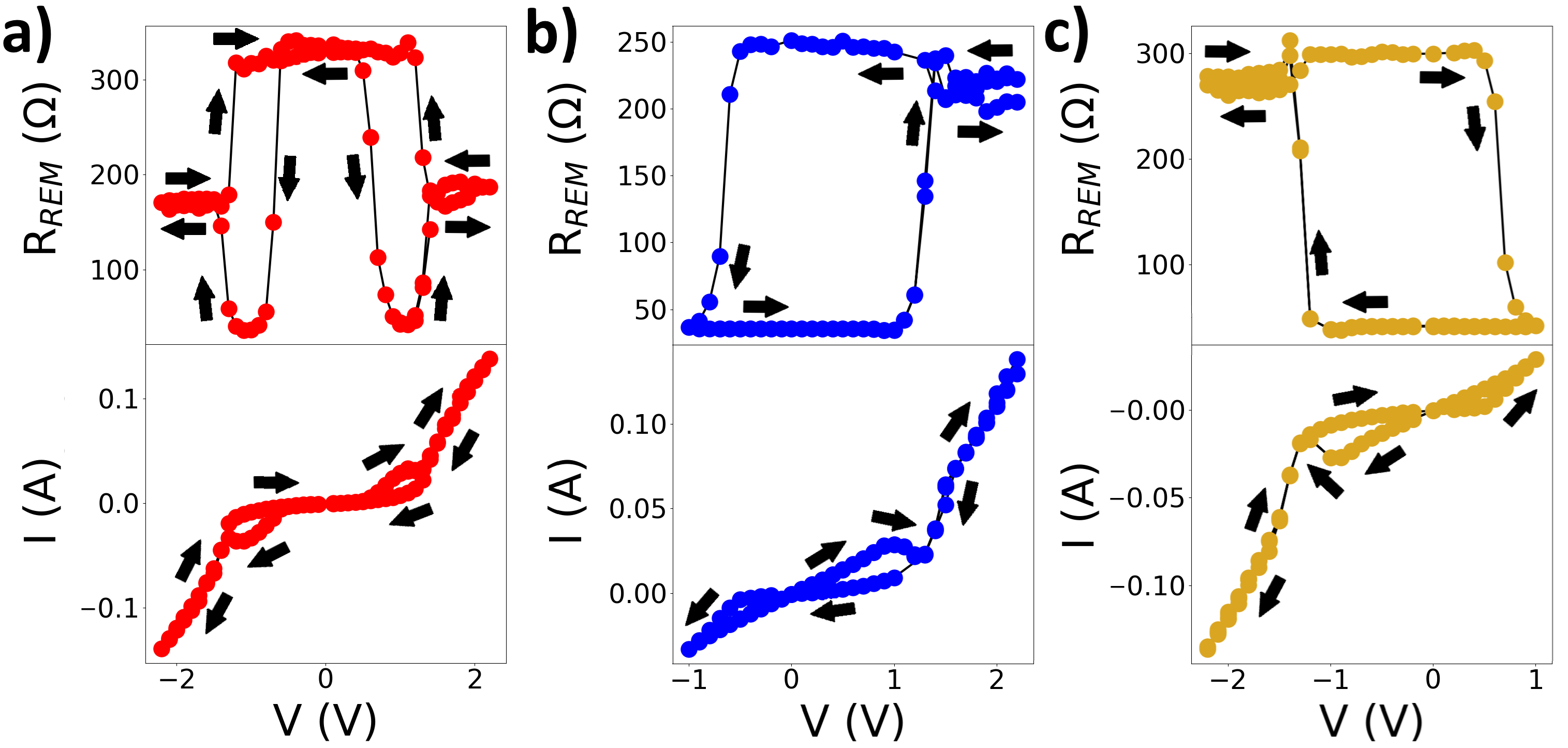}\caption{Electrical characterization of Ag-modified TaO$_x$ memristive devices. Remanent resistance, $R_{\mathrm{REM}}$ (top panels), and corresponding current--voltage ($I$--$V$) characteristics (bottom panels) measured under different programming protocols: (a) symmetric voltage cycling, revealing the characteristic highly symmetric table-with-legs (TWL) hysteresis switching loop with lateral ``wings'' and intermediate resistance states; (b) clockwise (CW) switching circulation; and (c) counter-clockwise (CCW) switching circulation. The TWL response reflects complementary SET and RESET processes occurring under opposite voltage polarities, while the similar behavior obtained for CW and CCW circulations highlights the high degree of polarity symmetry introduced by Ag incorporation. Black arrows indicate the direction of the applied-voltage sequence.}
\label{Fig3}
\end{figure}

Following AgNP incorporation, the electrical response changes qualitatively. As shown in FIG. \ref{Fig3}(a), the conventional CCW hysteresis is replaced by a highly symmetric `table-with-legs' (TWL) switching loop, indicative of complementary switching processes occurring under opposite voltage polarities \cite{Fer_2020,LMartir_2023}. Asymmetric voltage excursions selectively activate one of the two switching interfaces, transforming the symmetric TWL into nearly square hysteresis loops with opposite chiralities (FIGs. 3(b) and 3(c)). This interfacial selectivity is consistent with the complementary switching scenario previously established for TaO$_x$-based memristors \cite{Fer_2020}.

The evolution observed upon electrical stressing of the AgNPs-decorated devices is consistent with the incorporation of Ag-related species from the surface nanoparticles into the oxide. This process is accompanied by a pronounced overall reduction in device resistance, with the high-resistance state decreasing to approximately 300~$\Omega$ while the low-resistance state remains close to 40~$\Omega$.   A possible microscopic route for Ag incorporation into the oxide is a field-driven electrochemical process involving oxidation of metallic Ag, electromigration of Ag ions, and subsequent reduction. However, the present measurements do not directly resolve the charge state, migration pathway, or final microscopic configuration of Ag during operation, and this mechanism should therefore be regarded as a physically plausible scenario rather than an established description of the switching process.

A second remarkable feature of the HSL of FIG. 3(a) (symmetric stimulation) is the appearance of pronounced lateral extensions (`wings') emerging from the TWL at programming voltages close to $\pm1.5$~V. Rather than completing the RESET transition immediately after the resistance jump, the device evolves through intermediate resistance states that remain nearly constant over an extended voltage interval before the RESET process resumes. As shown in the Supplementary Material (FIG. S3), the `wings' persist for several hundred seconds and constitute a reproducible feature of the Ag-modified devices. In addition, the characteristic winged HSL was reproducibly observed across multiple Ag-modified devices, confirming that this response is not restricted to a single device (FIG. S1).

The TWL topology of the HSL, the high degree of polarity symmetry, and the emergence of metastable intermediate states indicate that Ag incorporation fundamentally modifies the switching dynamics of the TaO$_x$ memristors. As will be discussed later, these electrical characteristics are closely related to the electrically programmable synaptic depression dynamics observed under pulsed operation.

\begin{figure}[H] 
\centering
\includegraphics[width=1\linewidth]{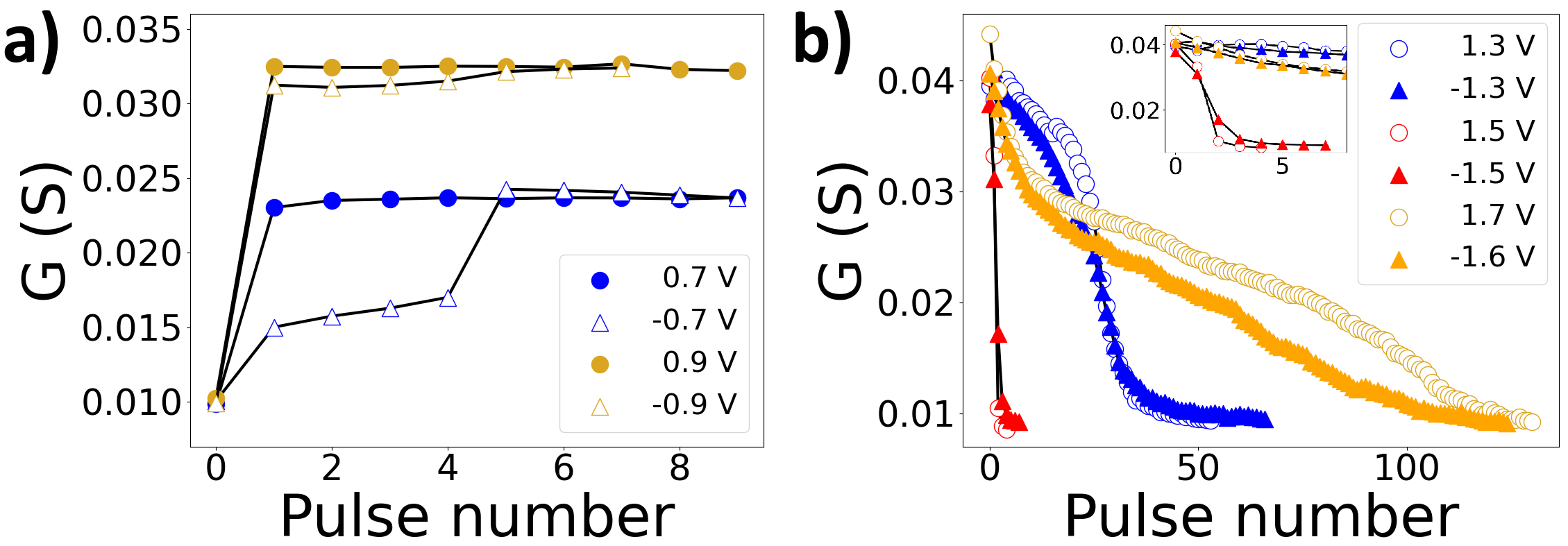}
\caption{Stimulus-dependent synaptic potentiation and depression in Ag-modified TaO$_x$ memristive devices. (a) Potentiation curves measured under different programming pulse amplitudes and for both switching polarities. In all cases, the conductance increases rapidly during the first few pulses and approaches a stable high-conductance state, with only a moderate dependence of the final conductance on pulse amplitude. (b) Depression curves measured under different programming amplitudes, revealing a strongly non-monotonic evolution of the conductance-update dynamics. Low-amplitude stimulation produces a gradual sigmoidal-like depression, intermediate amplitudes lead to an abrupt conductance decrease concentrated within the first few pulses, whereas higher amplitudes result in a broadly distributed conductance evolution extending over more than one hundred programming events. Filled symbols correspond to CCW switching circulation and open symbols to CW circulation. The close correspondence between both polarities highlights the high degree of symmetry of the Ag-modified devices.}
\label{Fig4}
\end{figure}

Potentiation and depression curves were measured for the Ag-modified memristor (FIG. \ref{Fig4}). Since the device exhibits both positive and negative SET and RESET transitions, the measurements were performed following the TWL cycle, namely, positive SET $\rightarrow$ positive RESET $\rightarrow$ negative SET $\rightarrow$ negative RESET. Potentiation (FIG. \ref{Fig4}(a)) remains qualitatively similar to that of the reference devices. In both switching polarities, the conductance increases rapidly during the first programming pulses and reaches stable high-conductance states after only a few pulses. A moderate dependence of the final conductance on pulse amplitude is observed, whereas the overall shape of the potentiation curves changes only slightly. The nearly identical potentiation characteristics obtained for both clockwise (CW) and CCW circulations further reflect the high degree of polarity symmetry of the Ag-modified switching response.

A markedly different behavior is observed during depression (FIG. \ref{Fig4}(b)). Unlike the reference devices, the Ag-modified memristors do not exhibit a single characteristic RESET dynamics. Instead, the conductance evolution changes markedly with programming amplitude, encompassing three characteristic operating regimes. At low pulse amplitudes (1.3~V), the conductance decreases progressively following a sigmoidal-like evolution over approximately 50 programming pulses. 
Increasing the programming amplitude to 1.5~V results in a dramatic acceleration of the depression process, with most of the conductance change taking place within the first 4--5 pulses. Remarkably, a further increase in pulse amplitude does not accelerate the RESET process. Instead, for pulse amplitudes of 1.6--1.7~V, the conductance evolves much more gradually over more than one hundred programming pulses without reaching complete saturation within the experimental window. Therefore, the depression dynamics exhibit a clearly non-monotonic dependence on programming voltage, with the fastest conductance evolution occurring at intermediate pulse amplitudes. 
Importantly, this behavior is reproduced for both switching polarities, indicating that it is not associated with a polarity-specific switching branch.
The ability to electrically tune the evolution of synaptic depression within the same physical device constitutes one of the central findings of this work. In contrast to conventional memristive synapses, where the conductance update dynamics are essentially fixed once the device architecture has been defined, AgNP incorporation enables a strongly non-monotonic modulation of the depression kinetics simply by adjusting the programming conditions.

\begin{figure}[H] 
\centering
\includegraphics[width=1\linewidth]{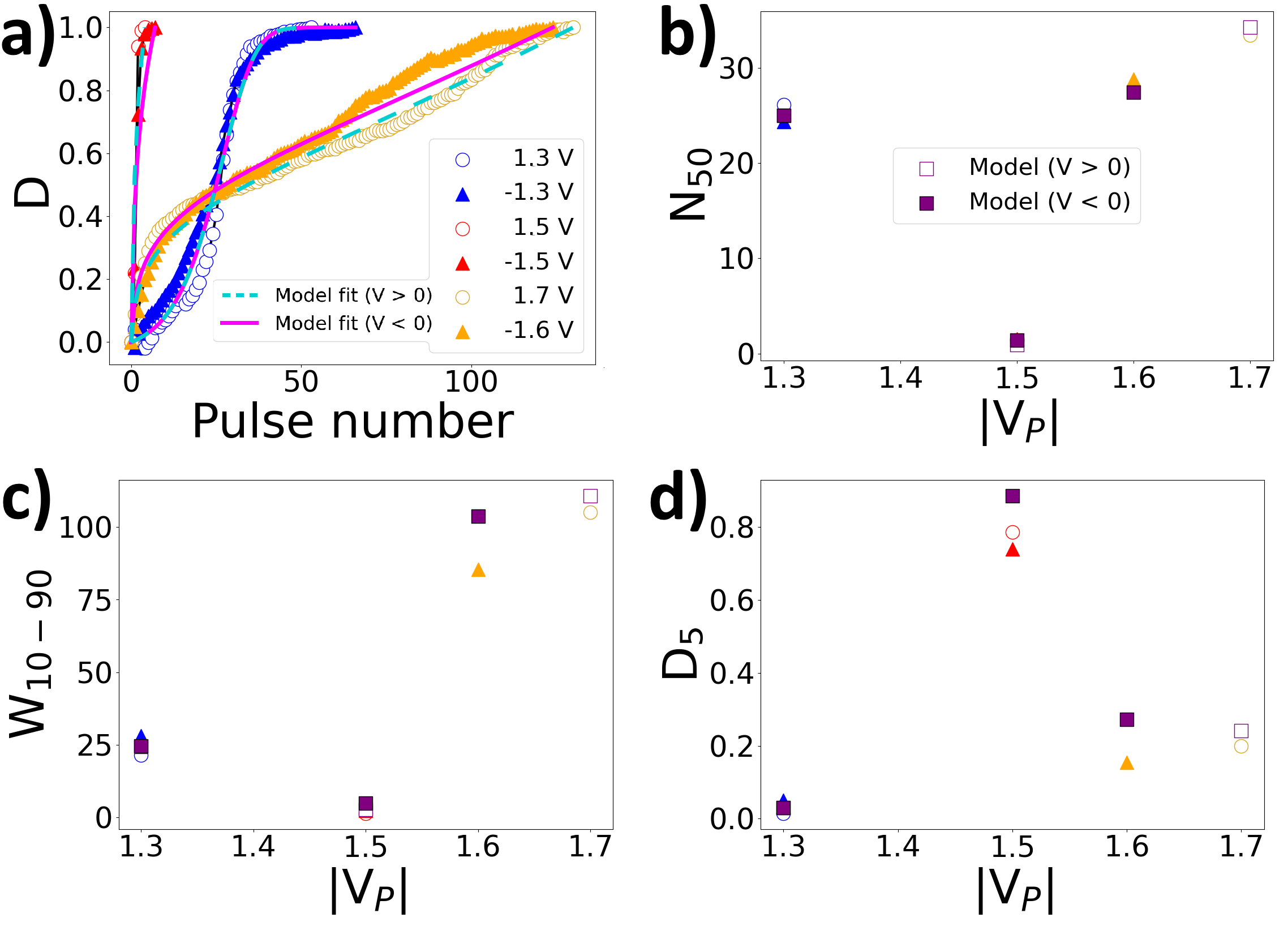}
\caption{Quantitative characterization of the stimulus-dependent depression dynamics in Ag-modified TaO$_x$ devices. (a) Normalized depression, $D(N)$, as a function of pulse number for different programming amplitudes and both voltage polarities. Symbols correspond to the experimental data, while the dashed and solid lines represent the global fits obtained with the minimal coupled-state model for positive and negative programming voltages, respectively (see Eqs(5)-(8)). (b) Characteristic pulse number $N_{50}$, defined by $D(N_{50})=0.5$, as a function of the programming-voltage magnitude $|V_P|$. (c) Transition width $W_{10-90}=N_{90}-N_{10}$, where $N_{10}$ and $N_{90}$ correspond to $D=0.1$ and $D=0.9$, respectively, quantifying the number of programming events over which most of the conductance change occurs. Square symbols correspond to values derived from the minimal coupled-state model. (d) Fraction of the total experimentally observed depression completed within the first five programming pulses, $D_5$ = $D(N=5)$. The simultaneous non-monotonic evolution of $N_{50}$, $W_{10-90}$, and $D_5$ reveals a crossover from gradual depression at low programming amplitudes to an abrupt response near $|V_P|=1.5$~V and to a broadly distributed conductance evolution at higher amplitudes. The close correspondence between positive and negative polarities indicates that the three regimes arise from a common underlying switching dynamics.}
\label{Fig5}
\end{figure}

To quantitatively assess the stimulus-dependent depression dynamics, we define the normalized depression as:


\begin{equation}
D(N)= \frac {1 - G(N)/G_0}{1-G_{end}/G_0},
\end{equation}



where $G_0$ is the initial conductance, $G(N)$ is the conductance after 
$N$ programming pulses, and $G_{\mathrm{end}}$ is the conductance measured 
at the end of the corresponding pulse sequence. Thus, $D=0$ and $D=1$ 
represent, respectively, the initial state and the total conductance change 
observed within each experimental window. This normalization allows the 
temporal evolution of the depression process to be compared independently 
of differences in the absolute conductance range. The normalized depression curves are shown in FIG.~\ref{Fig5}(a).

The non-monotonic evolution of D(N) with the voltage amplitude can be quantified by the characteristic pulse 
number $N_{50}$ at which \(D\) reaches 50$\%$ of the total normalized depression, defined through 

\begin{equation}
D(N_{50})=0.5.
\end{equation}

 As shown in FIG.~\ref{Fig5}(b), $N_{50}$ decreases from approximately 
25 pulses at $|V_{\mathrm{P}}|\simeq1.3$~V to approximately one pulse at 
$|V_{\mathrm{P}}|=1.5$~V, and subsequently increases again to approximately 
30 pulses for the highest programming amplitudes. Therefore, increasing the 
stimulus amplitude does not produce a monotonic acceleration of the RESET 
process. Instead, the fastest depression dynamics occurs within an 
intermediate voltage range. For $D\geq0.5$, no additional crossings between the depression curves are observed within the investigated voltage and pulse ranges.


Additional information on the temporal distribution of the conductance 
change is provided by the transition width,

\begin{equation}
W_{10-90}=N_{90}-N_{10},
\end{equation}

where $N_{10}$ and $N_{90}$ correspond to $D=0.1$ and $D=0.9$, 
respectively. FIG.~\ref{Fig5}(c) reveals an even stronger non-monotonic 
dependence on pulse amplitude. At low voltage, the depression extends over 
approximately 20--30 pulses, whereas at $|V_{\mathrm{P}}|=1.5$~V the 
transition is essentially completed within one or two programming events. 
At the highest amplitudes, $W_{10-90}$ increases dramatically to 
approximately 85--105 pulses. Thus, the high-voltage response cannot be 
described simply as a slower version of the intermediate-voltage RESET; 
rather, the conductance modification becomes distributed over a 
substantially broader programming window.

Finally, FIG.~\ref{Fig5}(d) shows the fraction of the experimentally observed depression completed after the first five programming pulses,
\begin{equation}
D_5 \equiv D(N=5)=\frac{G_0-G(5)}{G_0-G_{\mathrm{end}}}.
\end{equation}
Only a small fraction of the total conductance change occurs within the 
first five pulses at $|V_{\mathrm{P}}|\simeq1.3$~V, whereas approximately 
75\% of the depression is already completed at 
$|V_{\mathrm{P}}|=1.5$~V. At higher amplitudes, $D_5$ decreases again to 
approximately 15--20\%, consistently with the emergence of a more 
distributed conductance evolution.

Taken together, $N_{50}$, $W_{10-90}$, and $D_5$ provide independent and 
consistent evidence that the programming amplitude controls not only the 
magnitude or rate of the conductance change, but also its temporal 
distribution. In particular, an intermediate stimulus produces a strongly 
concentrated update, whereas both lower and higher amplitudes result in 
more gradual depression dynamics with distinct temporal profiles. The close 
correspondence between positive and negative stimulation further suggests 
that these regimes originate from a common underlying switching process.

\begin{figure}[H] 
\centering
\includegraphics[width=1\linewidth]{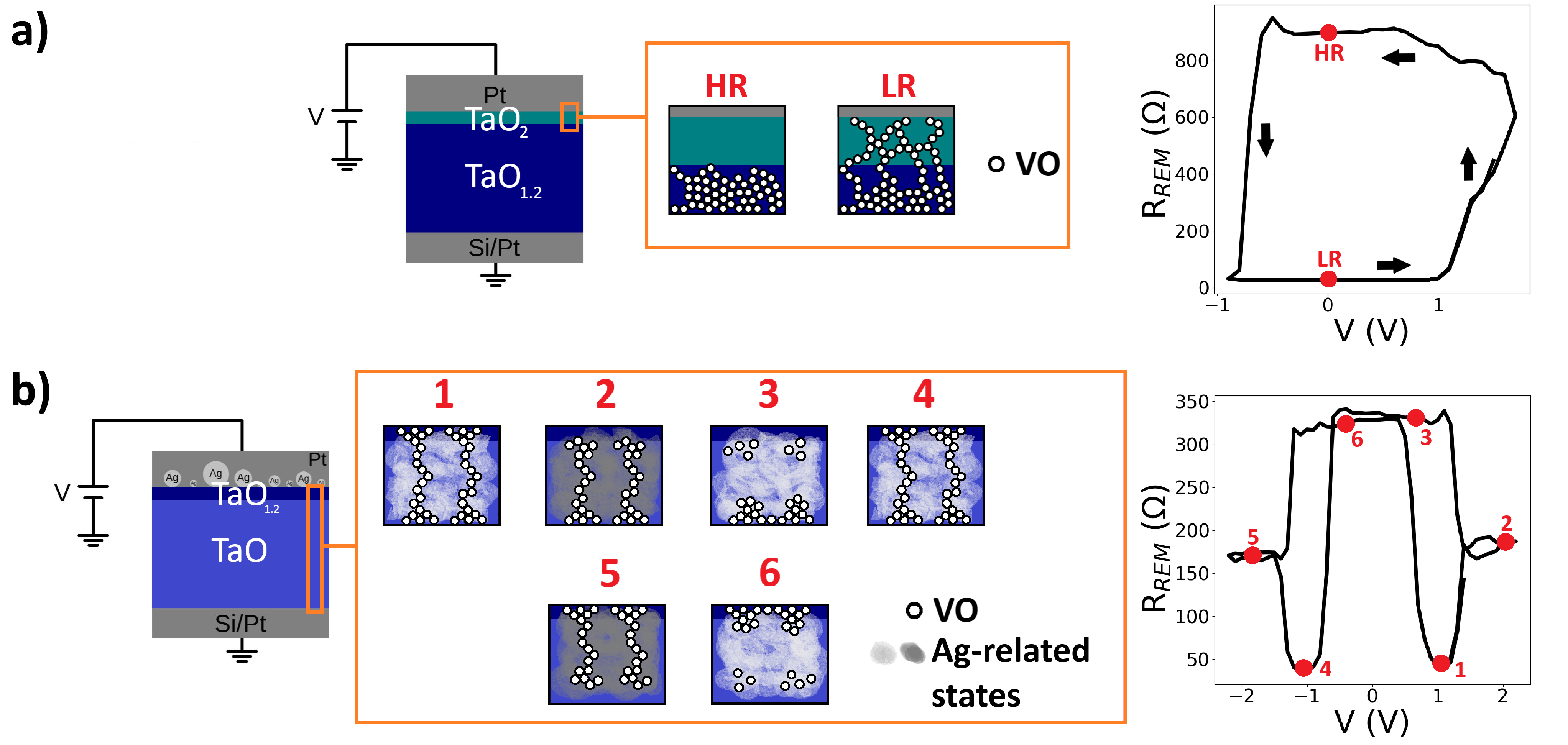}
\caption{Proposed physical scenarios for resistive switching in TaO$_x$ devices (a) without and (b) with AgNPs. (a) Reference device consisting of a more oxidized TaO$_2$ upper layer and a reduced TaO$_{1.2}$ bottom layer. The latter acts as an oxygen-vacancy reservoir, while the upper region constitutes the main active switching region. Under negative bias applied to the top electrode, positively charged oxygen vacancies (VO, white circles) migrate toward the upper interface, promoting the formation of a vacancy-rich conductive pathway and the low-resistance (LR) state. Positive bias promotes vacancy redistribution and partial disruption of this pathway, restoring the high-resistance (HR) state and yielding the conventional CCW hysteresis switching loop. (b) Ag-modified device, characterized by a more reduced TaO$_{1.2}$/TaO bilayer and a highly symmetric TWL hysteresis switching loop. The numbered sketches 1--6 correspond to the indicated points of the experimental HSL and schematically represent the coupled evolution of vacancy-rich conductive configurations and an additional Ag-related internal state during voltage cycling, with both metal/oxide interfaces participating in the switching process. White circles represent oxygen vacancies, while gray clouds denote the effective Ag-related degree of freedom. Their intensity schematically represents the field-dependent evolution of this internal state, with darker clouds denoting states more strongly coupled to oxygen vacancies. The gray clouds should not be interpreted as a direct microscopic determination of Ag position, charge state, or migration pathway, but as a physically consistent representation of the additional internal state invoked in the coupled-state model.}
\label{Fig6}
\end{figure}


Therefore, pulse amplitude acts as an external control parameter that 
selects the synaptic update dynamics within the same physical device. This 
stimulus-selectable response constitutes an additional degree of freedom 
beyond the conventional control of conductance levels and may be exploited 
to tailor memristive synapses to different learning requirements.

The simultaneous non-monotonic evolution of $N_{50}$,
$W_{10-90}$, and $D_5$ indicates that the observed crossover cannot be
described as a simple acceleration of a single field-driven RESET
process. In particular, the recovery of broad depression trajectories
at high programming amplitudes suggests the activation of an additional
dynamical degree of freedom. To test whether competition between two
field-activated processes is sufficient to reproduce the experimental
behavior, we introduce a minimal coupled-state model.

The first state variable, $x_N$, is an effective vacancy-related progress variable describing the evolution of the RESET trajectory after the $N$th programming pulse, while $a_N$ represents a slower Ag-related internal state. The latter is an effective dynamical variable and should not be identified with a directly measured Ag concentration. Their pulse-to-pulse evolution is described by

\begin{equation}
x_{N+1}
=
x_N+
\frac{A(V)}{1+\lambda a_N}\,
f(x_N),
\label{eq:model_x}
\end{equation}

and

\begin{equation}
a_{N+1}
=
a_N+
B(V)(1-a_N)-\gamma a_N,
\label{eq:model_a}
\end{equation}

where $\lambda\geq0$ quantifies the coupling between the two internal states and $\gamma\geq0$ represents relaxation of the Ag-related degree of freedom. The voltage-dependent rates are chosen to be strictly monotonic,

\begin{equation}
A(V)=
A_0
\exp\left[
\frac{|V|-V_{\mathrm{ref}}}{V_x}
\right],
\end{equation}

and

\begin{equation}
B(V)=
\frac{B_0}
{1+\exp[-(|V|-V_{\mathrm{th}})/\Delta V]}.
\end{equation}

Here, $f(x_N)$ accounts phenomenologically for the state dependence of the vacancy-related dynamics, while $V_x$, $V_{\mathrm{ref}}$, $V_{\mathrm{th}}$, and $\Delta V$ define the voltage dependence of the two elementary processes; their complete definitions are provided in the Supporting Information. For each programming voltage, the coupled equations are iterated pulse by pulse to obtain $x_N$ and $a_N$. The resulting $x_N$ trajectory is mapped onto the device conductance and subsequently normalized using the same definition of $D(N)$ employed for the experimental data. The complete model-to-observable mapping, initial conditions, and global fitting procedure are detailed in the Supporting Information.

Increasing $|V|$ therefore accelerates both elementary processes. At low programming amplitudes, the vacancy-related evolution is slow, resulting in gradual depression. At intermediate amplitudes, $A(V)$ becomes large while the Ag-related state remains weakly activated, producing a rapid RESET. At higher amplitudes, however, the progressive activation of $a_N$ reduces the effective vacancy-reorganization rate through the factor $1/(1+\lambda a_N)$, giving rise to a renewed, broadly distributed depression response. Thus, the gradual--abrupt--gradual crossover emerges from the competition between two individually monotonic field-activated processes rather than from an explicitly non-monotonic elementary rate.

A single global parameter set was fitted simultaneously to all measured depression trajectories and both voltage polarities. As shown in FIGs. ~5(a)-(d), the model reproduces the main features of the experimental crossover, including the pronounced minimum in $N_{50}$, the collapse and subsequent recovery of $W_{10-90}$, and the corresponding maximum in $D_5$. The agreement supports the coupled-state description as a minimal quantitative framework for the observed non-monotonic depression dynamics.

Based on the structural and electrical observations, and guided by the
coupled-state description above, we propose the physical scenario
schematically illustrated in FIG.~\ref{Fig6}. In devices without Ag, the increased oxidation of the upper layer results in a higher electrical resistance \cite{bao_2023}, thereby increasing the local electric field and favouring OV migration across this region. Meanwhile, the strongly reduced bottom TaO$_x$ layer acts as an oxygen-vacancy
reservoir for the more oxidized upper layer \cite{Fer_2020}. Under negative bias (SET), oxygen vacancies migrate toward the upper interface and form a vacancy-rich conductive pathway, producing the low-resistance state.
Positive bias promotes the reverse redistribution and partial disruption
of this pathway, restoring the high-resistance state. 


Ag incorporation substantially modifies this picture. The more uniform oxide stoichiometry and the highly symmetric table-with-legs loops suggest that both metal/oxide interfaces contribute to the switching process. We therefore propose that Ag primarily modifies the dynamics of the vacancy-mediated conductive network rather than introducing a fully independent switching mechanism such as conventional electrochemical metallization. A physically plausible route for this Ag-induced modification is the field-driven incorporation of Ag-related species into the oxide, potentially involving oxidation of metallic Ag, electromigration of Ag$^{+}$ ions, and subsequent reduction, as reported for Ag-containing memristive systems \cite{jeon_2024}. However, the present measurements do not directly resolve the charge state, migration pathway, or microscopic configuration of Ag during device operation.

Within this framework, the enhanced electric field in the vicinity of the Pt/TaO$_x$ interfaces promotes the redistribution of oxygen vacancies \cite{LealMartir_2026}, leading to the formation and partial disruption of vacancy-rich conductive pathways as the applied voltage is cycled (FIG.~\ref{Fig6}(b)). The presence of Ag introduces an additional internal degree of freedom that can modify this vacancy dynamics. In particular, sufficiently strong electrical stimulation may activate Ag-related defect configurations that transiently stabilize intermediate vacancy arrangements or modify the effective barriers for their reorganization. Such coupling may be further enhanced by the high local electric fields expected near partially disrupted conductive pathways \cite{chang_2017}. These metastable configurations provide a possible microscopic interpretation of the `wings' observed in the hysteresis loops and are consistent with the additional slow state required by the coupled-state model. Accordingly, the scheme in FIG.~\ref{Fig6} should be regarded as a physically consistent representation of coupled Ag-related and vacancy dynamics, rather than as a unique microscopic determination of Ag position, charge state, or motion.

To assess the functional implications of the electrically tunable
depression dynamics, we incorporated the different
conductance trajectories into a memristor-based neural-network
simulation. The calculations follow the hardware-oriented framework
described in Ref.~\cite{quinonez_2026}, in which synaptic weights are
encoded by memristive conductances and training is performed through
discrete conductance updates. In this framework, the distribution and
number of accessible conductance states determine the effective
resolution of the synaptic updates and can therefore strongly influence
learning performance.

To compare the functional impact of the three experimentally observed depression regimes, we generated synthetic depression trajectories reproducing their characteristic shapes while fixing the number of discrete conductance states to 100 in all cases. Following Ref.~\cite{quinonez_2026}, potentiation and depression were represented by discrete conductance sequences preserving the experimentally derived evolution of the conductance updates. The potentiation branch, network architecture, and training protocol
were kept identical across all simulations. The resulting differences in learning performance therefore reflect the combined effect of the temporal distribution and effective conductance excursion associated with each experimentally observed depression trajectory, rather than
differences in the nominal number of available update states.

\begin{figure}[H] 
\centering
\includegraphics[width=0.7\linewidth]{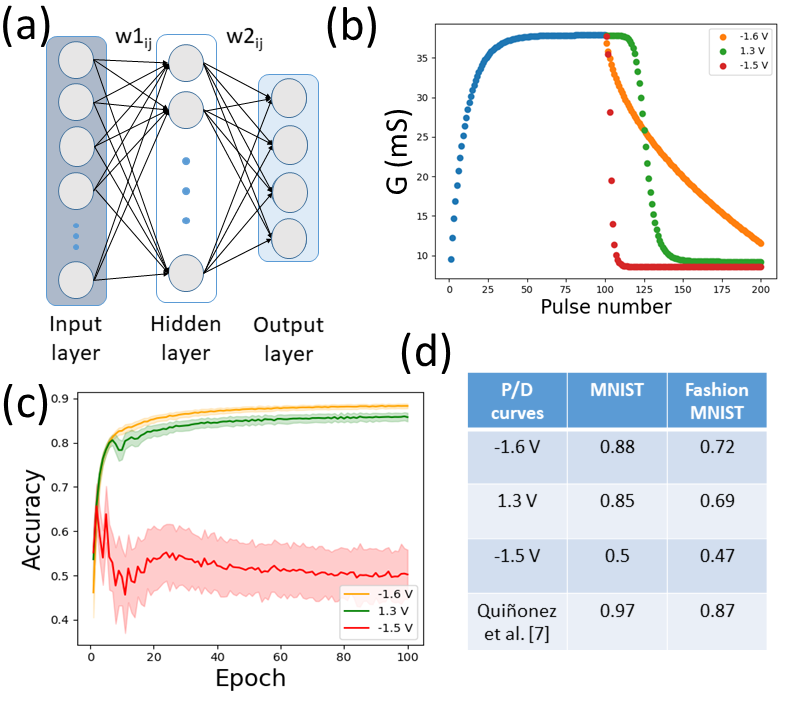}
\caption{Functional impact of the electrically selectable depression dynamics on neural-network learning. (a) Schematic representation of the single-hidden-layer multilayer perceptron (MLP) used in the simulations, with synaptic weights encoded by memristive conductances. The MLP consists of 784 input neurons, a single hidden layer with 100 neurons using ReLU activation, and 10 output neurons with softmax activation; training was performed using cross-entropy loss, mini-batches of 32 samples, 5-fold cross-validation, and 100 independent realizations of 100 epochs each. (b) Synthetic depression trajectories reproducing the three characteristic experimentally observed responses and associated with programming pulse amplitudes of $-1.6$~V, $1.3$~V, and $-1.5$~V, corresponding respectively to slowly evolving, sigmoidal, and abrupt depression regimes. All trajectories contain 100 discrete conductance states, ensuring the same nominal update-state count while preserving the characteristic conductance evolution of each experimentally observed depression regime. (c) Classification accuracy as a function of training epoch for the MNIST dataset using the three depression trajectories shown in (b), while keeping the network architecture, potentiation branch, and training protocol unchanged. Solid lines represent the mean accuracy and shaded regions indicate the dispersion between independent training realizations. (d) Summary of the final classification accuracies obtained for MNIST and Fashion-MNIST for the different depression trajectories, together with the reference values reported in Ref.~\cite{quinonez_2026} for more linear synthetic potentiation/depression characteristics.}
\label{fig7}
\end{figure}

FIG.~\ref{fig7} shows the resulting learning behavior for a
single-hidden-layer multilayer perceptron. For the MNIST classification
task, a pronounced dependence of network performance on the selected
depression trajectory is observed. The slowly evolving 
depression (-1.6 V) yields the highest accuracy among the investigated device
responses, approaching 88\%. The sigmoidal depression trajectory (1.3 V) also
supports robust learning, reaching an accuracy of approximately 85\%.
In marked contrast, the abrupt depression response (-1.5 V) strongly degrades
the learning process, resulting in an average accuracy of approximately
50\% together with a substantially larger dispersion between independent
training realizations ($\sigma \approx 0.06$).

The network performance was also evaluated using the more challenging Fashion-MNIST dataset. For the slow depression curve, the network reaches an
accuracy of approximately 72\%. This value is lower than the
approximately 87\% obtained in our previous simulations using
synthetic potentiation/depression curves with higher linearity
\cite{quinonez_2026}, as expected from higher nonidealities associated with the experimentally derived trajectory shape. Nevertheless,
the persistence of stable learning performance on Fashion-MNIST shows
that the gradual depression remains compatible with
training on a more demanding classification task.

The strong difference between the three MNIST results can be understood primarily in terms of the effective distribution and resolution of the synaptic updates. In the abrupt regime, a large fraction of the total conductance change is concentrated within only a few programming pulses, strongly restricting
the number of intermediate states that can be effectively accessed during depression. The resulting coarse weight updates hinder the progressive adjustment of the network weights and increase the sensitivity of the training process to individual update events, consistent with the large run-to-run dispersion. Conversely, the sigmoidal and, particularly, the slowly evolving depression trajectories distribute the conductance change over a larger number of programming events, providing finer effective weight updates and leading to substantially improved learning accuracy and reproducibility.

These simulations establish a direct functional connection between the
stimulus-controlled electrical response of the AgNP-containing devices
and their neuromorphic operation. Importantly, the applied programming
amplitude does not simply scale the amount or rate of depression.
Instead, it selects qualitatively different conductance-update
trajectories, ranging from abrupt to sigmoidal and slowly evolving
responses. The corresponding differences in neural-network performance
show that this electrically accessible dynamical degree of freedom has
direct consequences for learning. Within the conditions explored here,
the gradual depression regime provides the most favorable learning
behavior among the accessible responses, reaching
approximately 88\% accuracy for MNIST and 72\% for Fashion-MNIST.
 
\section{ DISCUSSION AND CONCLUSIONS}

Ag incorporation qualitatively changes the switching dynamics of studied TaO$_x$ memristors rather than simply modifying their resistance window. Reference devices exhibit conventional bipolar switching and rapidly saturating conductance updates, whereas Ag-modified devices display nearly symmetric table-with-legs hysteresis loops, long-lived intermediate resistance states, and a strongly non-monotonic depression response. Most notably, increasing the programming amplitude drives the same device through gradual, abrupt, and again broadly distributed depression regimes. The corresponding evolution of $N_{50}$, $W_{10-90}$, and $D_5$ shows that pulse amplitude controls not only the rate of conductance change, but also its temporal distribution. 

The coupled-state model provides a compact description of this behavior. The experimentally observed non-monotonicity is reproduced without introducing any non-monotonic elementary voltage dependence: both the vacancy-related rate $A(V)$ and the activation rate $B(V)$ of the second internal state increase monotonically with $|V|$. The gradual--abrupt--gradual crossover instead emerges from their competition. At low and intermediate amplitudes, vacancy-mediated RESET dominates, whereas at higher amplitudes activation of the second state reduces the effective vacancy-reorganization rate. A single global parameter set reproduces the normalized depression trajectories and the main evolution of $N_{50}$, $W_{10-90}$, and $D_5$, indicating that the anomalous voltage dependence can be understood as an emergent property of coupled internal dynamics rather than of a single switching channel.

This interpretation is consistent with the structural and electrical evidence. Ag is initially detected near the upper Pt/TaO$_x$ interface, while EDS in STEM-HAADF does not reveal extended metallic Ag clusters within the oxide. Together with the polarity symmetry, the table-with-legs response, and the long-lived intermediate states, these observations argue against a simple picture dominated by the formation and dissolution of a persistent metallic Ag filament. A more consistent scenario is that the dominant conductive network remains vacancy-mediated, while Ag incorporation modifies the local defect landscape and the kinetics of vacancy redistribution. Within this picture, the effective state variable $a_N$ may represent Ag-assisted changes in the local electric field, vacancy migration barriers, or stabilization of intermediate conductive configurations. The present measurements do not directly resolve Ag motion and therefore do not uniquely determine the microscopic nature of this state.

The winged switching topology was observed in all working Ag-modified devices, demonstrating that this non-conventional response is a robust and systematic feature of the Ag-containing structures rather than a device-specific anomaly. Notably, the onset of the wings occurs around |V| $\approx$ 1.5 V, coinciding with the strongest change in the depression kinetics and the transition toward the abrupt regime. This correlation supports a common Ag-modified dynamical origin for the electrically tunable depression.

The main functional consequence is that synaptic-update dynamics becomes an electrically selectable device property. Conventional pulse-amplitude control typically modifies the magnitude or rate of an otherwise similar conductance update. Recent work on W-modified BaTiO$_3$ memristors has similarly shown that defect engineering can reshape vacancy-mediated filament dynamics and enable systematic tuning of synaptic conductance through pulse amplitude, width, and number \cite{Ismail_2025}. In those devices, stronger or longer stimuli predominantly enhance the extent or rate of conductance modulation. In contrast, the response observed here is intrinsically non-monotonic, with programming amplitude giving access to qualitatively distinct depression trajectories that produce markedly different learning outcomes under otherwise identical network and training conditions.

These performance differences can be understood primarily in terms of the effective resolution of the synaptic updates. Abrupt depression compresses most of the conductance change into only a few programming events, producing coarse weight increments and limiting progressive weight refinement. More distributed trajectories instead spread the conductance change over a larger number of updates, providing finer effective weight increments and improving both accuracy and reproducibility. The relevant device variable is therefore not only the number or range of accessible conductance states, but also the trajectory through which those states are reached.

Taken together, these results establish a direct device--dynamics--function relationship: Ag incorporation reshapes the internal switching dynamics, electrical stimulation selects among the resulting update regimes, and those regimes determine learning performance. More broadly, the temporal structure of synaptic weight updates emerges not as a fixed by-product of device physics, but as an electrically programmable functional variable.

\section*{SUPPLEMENTARY MATERIAL}

The Supplementary Material includes additional electrical characterization of the Ag-modified TaO$_x$ devices, including device-to-device reproducibility and cycling stability of the characteristic table-with-legs hysteresis loops with `wings', measurements performed over extended programming-voltage ranges, and retention tests of the relevant resistance states. It also provides the full formulation of the minimal coupled-state model used to describe the non-monotonic depression dynamics, together with the conductance mapping, model assumptions, and the global fitting parameters used to reproduce the experimental depression trajectories.

\section*{ACKNOWLEDGMENTS}

D.R. acknowledges the HERON consortium (MSCA Staff Exchanges, Grant Agreement No. 101296294) for collaborative interactions related to this research.

\bibliographystyle{unsrt}
\bibliography{reference}

@article{Brink_2014,
    author = {ten Brink, Gert H. and Krishnan, Gopi and Kooi, Bart J. and Palasantzas, George},
    title = {Copper nanoparticle formation in a reducing gas environment},
    journal = {J. Appl. Phys.},
    volume = {116},
    number = {10},
    pages = {104302},
    year = {2014},
    month = {09},
    issn = {0021-8979},
    doi = {10.1063/1.4895483},
}

@article{Fer_2020,
	doi = {10.1088/1361-6528/ab6476},
	url = {https://doi.org/10.1088%2F1361-6528%2Fab6476},
	year = 2020,
	month = {},
	publisher = {{IOP} Publishing},
	volume = {31},
	number = {15},
	pages = {155204},
	author = {Ferreyra, Cristian and S{\'{a}}nchez, María José and Aguirre, Myriam and Acha, Carlos and  Bengi{\'{o}}, Silvina and Lecourt, Jerome and Lüders, Ulrike and Rubi, Diego},
	title = {Selective activation of memristive interfaces in \uppercase{T}a\uppercase{O}$_{x}$-based devices by controlling oxygen vacancies dynamics at the nanoscale},
	journal = {Nanotechnology}
}

@article{Gao_2017,
doi = {10.1088/1361-6528/aa6cd0},
url = {https://dx.doi.org/10.1088/1361-6528/aa6cd0},
year = {2017},
publisher = {IOP Publishing},
volume = {28},
number = {21},
pages = {215201},
author = {Gao, Leiwen and Li, Yanhuai and Li, Qin and Song, Zhongxiao and Ma, Fei},
title = {Enhanced resistive switching characteristics in \uppercase{A}l2\uppercase{O}$_3$ memory devices by embedded \uppercase{A}g nanoparticles},
journal = {Nanotechnology}
}

@book{IelminiWaser_2016,
  editor    = {Ielmini, Daniele and Waser, Rainer},
  title     = {Resistive Switching: From Fundamentals of Nanoionic Redox Processes to Memristive Device Applications},
  publisher = {Wiley-VCH},
  address   = {Weinheim, Germany},
  year      = {2016},
  doi       = {10.1002/9783527680870},
  isbn      = {9783527334179}
}

@article{Ismail_2025,
  author  = {Ismail, Muhammad and Na, Hyesung and Rasheed, Maria and Mahata, Chandreswar and Kim, Yoon and Kim, Sungjun},
  title   = {Synaptic metaplasticity and associative learning in low-power neuromorphic computing using {W}-diffused {BaTiO$_3$} memristors},
  journal = {Nano Energy},
  volume  = {142},
  pages   = {111276},
  year    = {2025},
  doi     = {10.1016/j.nanoen.2025.111276}
}

@article{LMartir_2023,
doi = {10.1088/1361-6528/aca597},
url = {https://dx.doi.org/10.1088/1361-6528/aca597},
year = {2022},
month = {dec},
publisher = {IOP Publishing},
volume = {34},
number = {9},
pages = {095202},
author = {Rodrigo {Leal Martir} and María José Sánchez and Myriam Aguirre and Walter Quiñonez and Cristian Ferreyra and Carlos Acha and Jerome Lecourt and Ulrike Lüders and Diego Rubi},
title = {Oxygen vacancy dynamics in \uppercase{P}t/\uppercase{T}i\uppercase{O}x/\uppercase{T}a\uppercase{O}y/\uppercase{P}t memristors: exchange with the environment and internal electromigration},
journal = {Nanotechnology}
}

@article{LealMartir_2026,
year = {2026},
publisher = {IOP Publishing},
volume = {59},
number = {20},
pages = {205302},
author = {Leal Martir, Rodrigo and van der Ree, Adrianus Julien Theodoor and Aguirre, Myriam H. and Palasantzas, George and Rubi, Diego and Sánchez, María José},
title = {Tuning the memristive response of \uppercase{T}a\uppercase{O}x- based devices with \uppercase{A}g nanoparticles},
journal = {J. Phys. D: Appl. Phys.}
}

@article{Lee2011,
  author={Lee, Myoung-Jae and Lee, Chang Bum and Lee, Dongsoo and Lee, Seung Ryul and Chang, Man and Hur, Ji Hyun and Kim, Young-Bae and Kim, Chang-Jung and Seo, David H. and Seo, Sunae and Chung, U-In and Yoo, In-Kyeong and Kim, Kinam},
  title={A Fast, High-Endurance and Scalable Non-Volatile Memory Device Made from Asymmetric \uppercase{T}a$_2$\uppercase{O}$_5-x$/\uppercase{T}a\uppercase{O}$_2-x$ Bilayer Structures},
  journal={Nat. Mater.},
  number={8},
  volume={10},
  pages={625--630},
  year={2011}
}

@article{Li2018,
  author={Li, Can and Hu, Miao and Li, Yunning and Jiang, Hao and Ge, Ning and Montgomery, Eric and Zhang, Jiaming and Song, Wenhao and Dávila, Noraica and Graves, Catherine E. and Li, Zhiyong and Strachan, John Paul and Lin, Peng and Wang, Zhongrui and Barnell, Mark and Wu, Qing and Williams, R. Stanley and Yang, J. Joshua and Xia, Qiangfei},
  title={Analogue Signal and Image Processing with Large Memristor Crossbars},
  journal={Nat. Electron.},
  volume={1},
  number={1},
  pages={52--59},
  year={2018}
}

@article{Ning_2021,
doi = {10.1088/1674-1056/abccb8},
year = {2021},
volume = {30},
number = {4},
pages = {047301},
author = {Ning, Yue and Lai, Yunfeng and Wan, Jiandong and Cheng, Shuying and Zheng, Qiao and Yu, Jinling},
title = {Implementation of synaptic learning rules by \uppercase{T}a\uppercase{O}$_x$ memristors embedded with silver nanoparticles},
journal = {Chin. Phys. B.}
}

@article{Prezioso2015,
  author={Prezioso, Mirko and Merrikh-Bayat, Farnood and Hoskins, Brian D. and Adam, Gina C. and Likharev, Konstantin K. and Strukov, Dimitri B.},
  title={Training and Operation of an Integrated Neuromorphic Network Based on Metal-Oxide Memristors},
  journal={Nature},
    volume = {521},
    number = {7550},
    pages = {61-64},
    year = {2015},
    doi = {10.1038/nature14441}
}

@article{Spring2020,
author = {Spring, Jonathan and Sediva, Eva and Hood, Zachary D. and Gonzalez-Rosillo, Juan Carlos and O'Leary, Willis and Kim, Kun Joong and Carrillo, Alfonso J. and Rupp, Jennifer L. M.},
title = {Toward Controlling Filament Size and Location for Resistive Switches via Nanoparticle Exsolution at Oxide Interfaces},
journal = {Small},
volume = {16},
number = {41},
pages = {2003224},
doi = {https://doi.org/10.1002/smll.202003224},
year = {2020}
}

@article{Strukov2008,
  author={Strukov, Dimitri B. and Snider, Gregory S. and Stewart, Duncan R. and Williams, R. Stanley},
  title={The Missing Memristor Found},
  journal={Nature},
  volume={453},
  pages={80--83},
  year={2008}
}

@article{Yang2013,
  author={Yang, J. Joshua and Strukov, Dimitri B. and Stewart, Duncan R.},
  title={Memristive Devices for Computing},
  journal={Nat. Nanotechnol.},
  volume={8},
  pages={13--24},
  year={2013}
}

@article{Yu2020,
  author={Yu, Shimeng},
  title={Neuro-Inspired Computing with Emerging Nonvolatile Memory},
  journal={Proceedings of the IEEE},
  volume={106},
  pages={260--285},
  year={2020}
}

@article{bao_2023,
title = {Electrical conductivity of \uppercase{T}a\uppercase{O}$_x$ as function of composition and temperature},
journal = {J. Non-Cryst. Solids.},
volume = {617},
pages = {122495},
year = {2023},
issn = {0022-3093},
doi = {https://doi.org/10.1016/j.jnoncrysol.2023.122495},
url = {https://www.sciencedirect.com/science/article/pii/S0022309323003617},
author = {Kefei Bao and Jingjia Meng and Jonathan D. Poplawsky and M. Skowronski}
}

@article{chang_2017,
author = {Chang, Chia-Fu and Chen, Jui-Yuan and Huang, Chun-Wei and Chiu, Chung-Hua and Lin, Ting-Yi and Yeh, Ping-Hung and Wu, Wen-Wei},
title = {Direct Observation of Dual-Filament Switching Behaviors in \uppercase{T}a2\uppercase{O}5-Based Memristors},
journal = {Small},
volume = {13},
number = {15},
pages = {1603116},
doi = {https://doi.org/10.1002/smll.201603116},
year = {2017}
}

@article{chen_2026,
author = {Chen, Shaochuan and Valov, Ilia},
title = {Enhanced Resistive Switching Uniformity in Tantalum Oxide Memristor Devices via Copper Implantation},
journal = {Adv. Electron. Mater.},
volume = {n/a},
number = {n/a},
pages = {e00002},
year = {2026},
doi = {https://doi.org/10.1002/aelm.202600002}
}

@article{covi_2015,
title = {Synaptic potentiation and depression in \uppercase{A}l:\uppercase{H}f\uppercase{O}2-based memristor},
journal = {Microelectron. Eng.},
volume = {147},
pages = {41-44},
year = {2015},
doi = {https://doi.org/10.1016/j.mee.2015.04.052},
author = {Covi, Erika and Brivio, Stefano and Fanciulli, Marco and Spiga, Sabrina}
}

@article{diao_2025,
    author = {Diao, Zhuo and Yamamoto, Ryohei and Meng, Zijie and Tohei, Tetsuya and Sakai, Akira},
    title = {Enhancing memristor multilevel resistance state with linearity potentiation via the feedforward pulse scheme},
    journal = {Nanoscale Horiz.},
    volume = {10},
    number = {4},
    pages = {780-790},
    year = {2025},
    doi = {10.1039/d4nh00623b}
}

@article{gao_2023,
title = {Tunable plasticity in functionalized honeycomb synaptic memristor for neurocomputing},
journal = {Mater. Today Phys.},
volume = {30},
pages = {100947},
year = {2023},
doi = {https://doi.org/10.1016/j.mtphys.2022.100947},
author = {Qin Gao and Jiangshun Huang and Juan Gao and Xueli Geng and Yuhang Ji and Haoze Li and Guoxing Wang and Bo Liang and Mei Wang and Zhisong Xiao and Ying Zhu and Paul K. Chu and Anping Huang}
}

@article{jana_2025,
title = {Mitigation of memory state variability in \uppercase{S}i\uppercase{O}$_2$ memristors by \uppercase{C}u nanoparticles incorporation},
journal = {J. Alloys Compd.},
volume = {1021},
pages = {179690},
year = {2025},
issn = {0925-8388},
doi = {https://doi.org/10.1016/j.jallcom.2025.179690},
author = {Biswajit Jana and Kritika Ghosh and Ayan {Roy Chaudhuri}}
}

@article{jeon_2024,
    author = {Jeon, Yu-Rim and Akinwande, Deji and Choi, Changhwan},
    title = {Volatile threshold switching and synaptic properties controlled by \uppercase{A}g diffusion using Schottky defects},
    journal = {Nanoscale Horiz.},
    volume = {9},
    number = {5},
    pages = {853-862},
    year = {2024},
    doi = {10.1039/d3nh00571b},
}

@article{lee_2011,
author = {Lee, MyoungJae and Lee, Chang Bum and Lee, Dongsoo and Lee, Seung Ryul and  Chang, Man and  Hur, Ji Hyun  and  Kim, Young-Bae and  Kim, Chang-Jung and Seo, David H. and Seo, Sunae and Chung, U-In and Yoo, In-Kyeong and Kim, Kinam },
title = {A fast, high-endurance and scalable non-volatile memory device made from asymmetric \uppercase{T}a${_2}$\uppercase{O}$_{5-x}$/\uppercase{T}a\uppercase{O}$_{2-x}$ bilayer structures},
journal = {Nat. Mater.},
volume = {10},
pages = {625},
year = {2011},
L3 = {https://www.nature.com/articles/nmat3070#supplementary-information},
URL = {https://doi.org/10.1038/nmat3070}
}

@article{liu_2010,
author = {Liu, Qi and Long, Shibing and Lv, Hangbing and Wang, Wei and Niu, Jiebin and Huo, Zongliang and Chen, Junning and Liu, Ming},
title = {Controllable Growth of Nanoscale Conductive Filaments in Solid-Electrolyte-Based \uppercase{R}e\uppercase{RAM} by Using a Metal Nanocrystal Covered Bottom Electrode},
journal = {ACS Nano},
volume = {4},
number = {10},
pages = {6162-6168},
doi ={10.1021/nn1017582},
year = {2010}
}

@article{meng_2020,
    author = {Qi, Meng and Cao, Shuo and Yang, Liu and You, Qi and Shi, Libin and Wu, Zhiying},
    title = {Uniform multilevel switching of graphene oxide-based \uppercase{R}\uppercase{R}\uppercase{A}\uppercase{M} achieved by embedding with gold nanoparticles for image pattern recognition},
    journal = {Appl. Phys. Lett.},
    volume = {116},
    number = {16},
    pages = {163503},
    year = {2020},
    doi = {10.1063/5.0003696},
}

@article{milano_2022,
    author = {Milano, Gianluca and Pedretti, Giacomo and Montano, Kevin and Ricci, Saverio and Hashemkhani, Shahin and Boarino, Luca and Ielmini, Daniele and Ricciardi, Carlo},
    title = "{In materia reservoir computing with a fully memristive architecture based on self-organizing nanowire networks}",
    journal = {Nat. Mater.},
    volume = {21},
    number = {2},
    pages = {195-202},
    year = {2022},
    month = {02},
    doi = {10.1038/s41563-021-01099-9}
}

@article{park_2015,
	Author = {Park, Tae Hyung and Song, Seul Ji and Kim, Hae Jin and  Kim, Soo Gil and Chung, Suock and Kim, Beom Yong and  Lee, Kee Jeung and Kim, Kyung Min and Choi, Byung Joon
 and  Hwang, Cheol Seong},
	Date = {2015/11/03/online},
	Day = {17},
	Journal = {Sci. Rep.},
	L3 = { 10.1038/srep15965 },
	M3 = {Article},
	Month = {},
	Pages = {15965},
	Publisher = {The Author(s) SN  -},
	Title = {Thickness effect of ultra-thin \uppercase{T}a$_2$\uppercase{O}$_5$ resistance switching layer in 28 nm-diameter memory cell},
	Ty = {JOUR},
	Url = { https://doi.org/10.1038/srep15965},
	Volume = {5},
	Year = {2015}}

@article{quinonez_2026,
    author = {Quiñonez, Walter and Sánchez, María José and Rubi, Diego},
    title = {Modeling memristor-based neural networks with \uppercase{M}anhattan update: \uppercase{T}rade-offs in learning performance and energy consumption},
    journal = {APL Electron. Devices},
    volume = {2},
    number = {2},
    pages = {026102},
    year = {2026},
    doi = {10.1063/5.0310714}
}

@article{roz_2010,
  title = {Mechanism for bipolar resistive switching in transition-metal oxides},
  author = {Rozenberg, Marcelo J. and S\'anchez, María José and Weht, Rubén and Acha, Carlos and Gomez-Marlasca, Fernando and Levy, Pablo},
  journal = {Phys. Rev. B},
  volume = {81},
  issue = {11},
  pages = {115101},
  numpages = {5},
  year = {2010},
  month = {},
  publisher = {American Physical Society},
  doi = {10.1103/PhysRevB.81.115101},
  url = {https://link.aps.org/doi/10.1103/PhysRevB.81.115101}
}

@article{song_2023,
title = {Doping modulated ion hopping in tantalum oxide based resistive switching memory for linear and stable switching dynamics},
journal = {Appl. Surf. Sci},
volume = {631},
pages = {157356},
year = {2023},
doi = {https://doi.org/10.1016/j.apsusc.2023.157356},
author = {Young-Woong Song and Yun-Hee Chang and Jaeho Choi and Min-Kyu Song and Jeong Hyun Yoon and Sein Lee and Se-Yeon Jung and Wooho Ham and Jeong-Min Park and Hyun-Suk Kim and Jang-Yeon Kwon}
}

@article{sudheer_2023,
title = {Linearly potentiated synaptic weight modulation at nanoscale in a highly stable two-terminal memristor},
journal = {Appl. Surf. Sci.},
volume = {610},
pages = {155411},
year = {2023},
doi = {https://doi.org/10.1016/j.apsusc.2022.155411},
author = { Sudheer and Rupam Mandal and Dilruba Hasina and Alapan Dutta and Safiul {Alam Mollick} and Aparajita Mandal and Tapobrata Som}
}

@article{vanderRee_2024,
  author    = {van der Ree, Adrianus Julien Theodoor and 
               Ahmadi, Majid and 
               Ten Brink, Gert H. and 
               Kooi, Bart J. and 
               Palasantzas, George},
  title     = {Stable Millivolt Range Resistive Switching in Percolating Molybdenum Nanoparticle Networks},
  journal   = {ACS Appl. Mater. Interfaces},
  year      = {2024},
  volume    = {16},
  number    = {47},
  pages     = {65157--65164},
  doi       = {10.1021/acsami.4c12051},
  publisher = {American Chemical Society}
}

@article{xiaobing_2018,
author = {Yan, Xiaobing and Zhao, Jianhui and Liu, Sen and Zhou, Zhenyu and Liu, Qi and Chen, Jingsheng and Liu, Xiang Yang},
title = {Memristor with \uppercase{A}g-\uppercase{C}luster-Doped \uppercase{T}i\uppercase{O}2 Films as Artificial Synapse for Neuroinspired Computing},
journal = {Adv. Funct. Mater.},
volume = {28},
number = {1},
pages = {1705320},
doi = {https://doi.org/10.1002/adfm.201705320},
year = {2018}
}

@article{yang_2012,
author = {Yang,Yuchao  and Sheridan,Patrick  and Lu,Wei },
title = {Complementary resistive switching in tantalum oxide-based resistive memory devices},
journal = {Appl. Phys. Lett.},
volume = {100},
number = {20},
pages = {203112},
year = {2012},
doi = {10.1063/1.4719198},
}

@article{yang_2014,
    author = {Yang, Yuchao and Gao, Peng and Li, Linze and Pan, Xiaoqing and Tappertzhofen, Stefan and Choi, ShinHyun and Waser, Rainer and Valov, Ilia and Lu, Wei D.},
    title = {Electrochemical dynamics of nanoscale metallic inclusions in dielectrics},
    journal = {Nat. Commun.},
    volume = {5},
    number = {1},
    pages = {4232},
    year = {2014},
    doi = {10.1038/ncomms5232}
}

@article{yue_2023,
title = {Improving resistive switching effect by embedding gold nanoparticles into ferroelectric thin films},
journal = {J. Alloys Compd.},
volume = {968},
pages = {171832},
year = {2023},
issn = {0925-8388},
doi = {https://doi.org/10.1016/j.jallcom.2023.171832},
author = {Zhi Yun Yue and Zhi Dong Zhang and Zhan Jie Wang}
}

@article{zhou_2022,
author = {Zhou, Guangdong and Wang, Zhongrui and Sun, Bai and Zhou, Feichi and Sun, Linfeng and Zhao, Hongbin and Hu, Xiaofang and Peng, Xiaoyan and Yan, Jia and Wang, Huamin and Wang, Wenhua and Li, Jie and Yan, Bingtao and Kuang, Dalong and Wang, Yuchen and Wang, Lidan and Duan, Shukai},
title = {Volatile and Nonvolatile Memristive Devices for Neuromorphic Computing},
journal = {Adv. Electron. Mater.},
volume = {8},
number = {7},
pages = {2101127},
doi = {https://doi.org/10.1002/aelm.202101127},
url = {https://onlinelibrary.wiley.com/doi/abs/10.1002/aelm.202101127},
eprint = {https://onlinelibrary.wiley.com/doi/pdf/10.1002/aelm.202101127},
year = {2022}
}

@article{zhu_2017,
title = {Enhanced stability of filament-type resistive switching by interface engineering},
journal = {Sci. Rep.},
volume = {7},
pages = {43664},
year = {2017},
number = {1},
doi = {10.1038/srep43664},
author = {Zhu, Ying Bin and Zheng, K. and Wu, Xing and Ang,  Lay-Kee}
}

@article{zhu_2024,
author = {Zhu, Mingmin and Yu, Zhendi and Hu, Gao and Yu, Kai and Jiang, Yulong and Wang, Jiawei and Dong, Wenjing and Guo, Jinming and Qiu, Yang and Yu, Guoliang and Zhou, Hao-Miao},
title = {A \uppercase{T}a\uppercase{O}$_x$/\uppercase{T}i\uppercase{O}$_y$ Bilayer Memristor with Enhanced Synaptic Features for Neuromorphic Computing},
journal = {Adv. Electron. Mater.},
volume = {10},
number = {8},
pages = {2400008},
doi = {https://doi.org/10.1002/aelm.202400008},
year = {2024}
}

@article{zhu_2025,
title = {Doping engineering in tantalum oxide-based \uppercase{R}\uppercase{R}\uppercase{A}\uppercase{M} with enhanced resistive switching behaviors and synaptic features for neuromorphic applications},
journal = {Mater. Des.},
volume = {260},
pages = {115199},
year = {2025},
doi = {https://doi.org/10.1016/j.matdes.2025.115199},
author = {Mingmin Zhu and Hui Ouyang and Zhendi Yu and Yu Du and Guangxiao Song and Wenjing Dong and Jiawei Wang and Yang Qiu and Guoliang Yu and Yan Li and Xufeng Jing and Haibin Zhu and Hao-Miao Zhou}
}

\end{document}